\documentclass{article}
\providecommand{\keywords}[1]
{
  \small	
  \textbf{\textit{Keywords---}} #1
}
\usepackage{geometry}
\usepackage{amsmath,amsfonts}
\usepackage{array}
\usepackage{xcolor} 
\usepackage{textcomp}
\usepackage{stfloats}
\usepackage{url}
\usepackage{verbatim}
\usepackage{graphicx}
\usepackage{cite}
\usepackage{subfig}
\usepackage{adjustbox}
\usepackage{mathtools}
\usepackage{float}
\usepackage{setspace}
\usepackage{fancyvrb}
\usepackage[ruled]{algorithm2e}
\usepackage{syntax}
\usepackage{bm}

\newcommand{\myttuline}[1]{\texttt{\underline{#1}}}
\newcommand{\myboldtt}[1]{\footnotesize{\textsc{\textbf{$\bm{<}$#1$\bm{>}$}}}}
\newcommand{\genia}{\textit{GenIA\textsuperscript{3}}}

\newcommand{\egenia}{\textit{e-GenIA\textsuperscript{3}}}

\title{
{\egenia}: An AgentSpeak extension for empathic agents
}          

\author{Author Surname$^{1}$, Someone Else$^{2}$  \\
        \small $^{1}$University A \\
        \small $^{2}$University B \\
}

\author{Joaquin Taverner$^{1,*}$, Emilio Vivancos$^{1}$, and Vicente Botti$^{1}$\\
\small $^{1}$Valencian Research Institute for Artificial Intelligence, Universitat Polit\`ecnica de Val\`encia, Spain\\
\small $^{*}$ Corresponding author: Joaquin Taverner (joataap@dsic.upv.es)\\}

\begin{document}
\maketitle

\begin{abstract}
In this paper, we present {\egenia} an extension of AgentSpeak to provide support to the development of empathic agents. The new extension modifies the agent's reasoning processes to select plans according to the analyzed event and the affective state and personality of the agent. In addition, our proposal allows a software agent to simulate the distinction between self and other agents through two different event appraisal processes: the empathic appraisal process, for eliciting emotions as a response to other agents emotions, and the regular affective appraisal process for other non-empathic affective events. The empathic regulation process adapts the elicited empathic emotion based on intrapersonal factors (e.g., the agent's personality and affective memory) and interpersonal characteristics of the agent (e.g., the affective link between the agents). The use of a memory of past events and their corresponding elicited emotions allows the maintaining of an affective link to support long-term empathic interaction between agents.
\end{abstract}

\keywords{Empathy, affective computing, empathic regulation, appraisal, multi-agent systems}

\section{Introduction}\label{sec:introduction}

Computer systems that are oriented to the paradigm of human-machine interaction are becoming more relevant in society. Systems of this type are progressively becoming more complex and require higher level abstractions and metaphors to describe capabilities and characteristics that cannot be explained by classical lower level specifications. 
Endowing software agents with affective abilities, particularly empathic abilities, is one of the metaphors that will make it possible to improve the simulation of human affective behavior and to transform the human-machine interaction paradigm by making it more human-oriented. 
Empathy is a concept that has evolved over the years, whose meaning refers to a wide range of affective and cognitive competencies that are fundamental in the development of the human being as a social being \cite{hoffman1984interaction}. These competencies allow one ``to put oneself in others' shoes'' and to understand and share their feelings and thoughts, which generates a series of pro-social behaviours aimed at improving their well-being \cite{davis2018empathy}. 
In any empathic interaction, we can distinguish two actors: the \textit{target} actor, who is the person or agent that suffers the effect of an action or situation, and the \textit{observer}, who is the person (or agent) that perceives that action or situation and feels empathy for the \textit{target} person.

Different theoretical approaches that try to explain the cognitive processes related to empathy have been proposed in psychology, sociology, philosophy, ethology, and neuroscience. The most recent theoretical approaches provide a conceptualization of empathy from a perspective derived from appraisal theories that relate the emergence of emotions to a cognitive appraisal of an event \cite{lazarus1991emotion,frijda1989relations}. Under this perspective, empathy arises from the appraisal of the perception of a situation or an emotion in others \cite{wondra2015appraisal,heyes2018empathy}. Moreover, according to several authors, empathy is affected by a regulation process that modulates the empathic response to an event according to different factors, including intrapersonal (e.g., personality) and interpersonal (e.g., affective bond or social tie with the \textit{target}) factors \cite{davis2018empathy,decety2004functional,clark2019feel}. 

For years, the inherent characteristics of human affective behavior have been the subject of research in the field of affective computing \cite{picard1997affective}. In this field, there has been a growing trend towards the development of models to simulate empathy \cite{paiva2017empathy,yalccin2019modeling}. Most of these models use the agent-oriented paradigm. Empathic agents have been shown in multiple experiments to improve the user experience in human-machine interactions \cite{rodrigues2015process,ochs2012formal,obaid2018endowing}. In general, these models are designed to be implemented ad hoc making the agent specification programmer and domain dependent. 

This paper presents {\egenia}, an extension of the syntax, semantics, and the reasoning cycle of the well-known agent-oriented language AgentSpeak \cite{rao1996agentspeak} to support the development of agents with empathic abilities implemented on the {\genia} architecture \cite{alfonso2017toward}. Our model is based on recent empathic appraisal theories to elicit empathic emotions when a software agent perceives or is aware of an emotion or a situation in other agents or humans. In addition, our empathic agent model has an \textit{emotion regulation} process that adapts emotions to different intrapersonal and interpersonal factors. 

The rest of this paper is organized as follows. In Section~\ref{relatedWork}, theoretical frameworks supporting affective states and empathy along with a discussion of proposals made in affective computing to simulate affective and empathic capabilities are presented. Section~\ref{section:genia} describes the {\genia} architecture. Section~\ref{proposal} introduces {\egenia}, an AgentSpeak extension to facilitate the development of agents with empathic abilities. In Section~\ref{defaultDsign} a default design for all the processes described in Section~\ref{proposal} is presented. Finally, the main conclusions and some future works are presented in Section~\ref{conclusion}.

\section{Related work}\label{relatedWork}
For years, the area of affective computing \cite{picard1997affective} has been working to design models to understand and simulate non-rational behaviors such as affective behaviors \cite{marsella2009ema,paiva2017empathy,taverner2020multidimensional}. These models generally include different constructs such as affective states, personality, or empathy that come from different branches of psychology, ethology, philosophy, and sociology. Over the years, different theorist have tried to provide an answer to the phenomenon of affective states, which involves constructs such as emotions and moods.  
One of the most recognized theories is the appraisal theory \cite{lazarus1991emotion,frijda1989relations}, according to which, when a stimulus is received, an appraisal process occurs, resulting in an emotion. Emotion is generally considered to be a quick response to certain stimuli with short duration. Appraisal theory is based on the existence of different variables known as \textit{Appraisal Variables}. Different authors differ in the number and type of \textit{Appraisal Variables} involved in the emotion generation process. For example, K.R. Scherer \cite{scherer1993studying} proposed a model in which twenty-two \textit{Appraisal Variables} were defined, while A. Ortony, G. Clore, and A. Collins \cite{ortony1990cognitive} proposed a more simplified model, known as OCC, with only eight \textit{Appraisal Variables}. However, appraisal theory is not the only theory that has been developed to explain the phenomenon of emotion. \textit{Basic emotion} theories relate events to a limited number of emotions. For example, the basic theory of emotions proposed by P. Ekman \cite{ekman1992argument} uses six basic emotions: \textit{Happiness}, \textit{Surprise}, \textit{Fear}, \textit{Anger}, \textit{Disgust}, and \textit{Sadness}. In contrast to these basic emotion theories, the constructivist theories advocate for a more universal concept of emotion. Constructivism holds the existence of an unlimited number of emotions which can present cross-cultural differences \cite{russell1989cross}. One of the most recognized constructivists is J.A. Russell who proposed the Circumplex Model of Affect \cite{russell1980circumplex}. According to Russell, emotions can be expressed in a two-dimensional space that is composed of the dimensions of \textit{Pleasure} and \textit{Arousal}. This theory is in line with the findings of neuroscience made in recent years \cite{posner2005circumplex}. More recent theories suggest the need to use models with more dimensions, such as \textit{Dominance} or \textit{Novelty}, to better explain the differences between emotions  \cite{mehrabian1996pleasure,scherer2019towards}. 

On the other hand, it is generally accepted that, in contrast to emotions, moods are unfocused and diffuse affective states that have a longer duration (from minutes to days) and a lower intensity than emotions \cite{beedie2005distinctions}. The variation in mood duration is sometimes attributed to the time the mood spends in transitioning to a neutral state or equilibrium state \cite{hilpert2020can}. Both the equilibrium state and the speed of transition may vary depending on the individual's affective characteristics. Moreover, it should be considered that both emotions and mood are influenced by personality \cite{rusting1997extraversion}. Empirical evidence suggests that personality is involved in emotion regulation processes 
making individuals more or less prone to certain emotions or moods \cite{wang2009neuroticism,ekkekakis2012affect}. One of the most recurrent theories of personality is the OCEAN model \cite{mccrae1992introduction}. This model defines personality through five dimensions: \textit{Openness}, \textit{Conscientiousness}, \textit{Extraversion}, \textit{Agreeableness}, and \textit{Neuroticism}. These five dimensions have an effect on the affective states that can be elicited in an individual. Apparently, the most evident relationships arise when relating positive affective states to \textit{Extraversion} and when relating negative affective states to \textit{Neuroticism} \cite{derryberry1994temperament}. 

Finally, empathy is a construct that is used in different domains, such as psychology, ethology, sociology, or philosophy, to describe a variety of psychological attitudes that enable the development of social individuals, encompassing a large number of emotional, ethical, moral, and social aspects \cite{packard2021we,cuff2016empathy}. In general terms, empathy is an ability that allows humans to understand and feel the affective state of others, resulting in behavior directed toward mutual understanding. Several authors consider empathy to be a fundamental ability that allows the establishment and maintenance of social bonds \cite{segal2018social,hoffman2008empathy,riess2017science}. Empathy also plays a fundamental role in our society, affecting both morality and mutual understanding and promoting relationships and collaborations through the exchange of experiences, needs, and desires \cite{davis2018empathy,marsh2018neuroscience}. 

In its earlier definition, empathy was described as an innate instinct that produces a self-awareness in the experience and awareness of the \textit{target} in the \textit{observer} without any perspective-taking, associative, or cognitive process \cite{stueber2013empathy,ganczarek2018einfuhlung}. This concept evolved over the years in different theories until the discovery of mirror neurons \cite{rizzolatti1996premotor}, which established a relationship between perception and action allowing a greater understanding of how imitation and empathy are produced in the brain and are related to emotion experiences \cite{lamm2019imaging,ferrari2017two}. Based on this relationship between perception and action, S.D. Preston and F. De Waal \cite{preston2002empathy} proposed a theory of empathy known as the \textit{Perception Action Model} (PAM). According to the PAM, when the \textit{observer} perceives an emotion in the \textit{target}, he/she can experience the same emotion automatically and not consciously, producing a matching of mental states between the \textit{observer} and the \textit{target} \cite{preston2007perception}. The PAM is used as a basis for explaining the evolution of higher-level cognitive processes related to cognitive empathy. For example, De Waal proposes the \textit{The Russian Doll} model \cite{de2007russian}. This model uses the analogy of a Russian doll, stratifying the empathic processes in different layers. The inner layers represent the most primitive mechanisms of empathy related to the PAM such as mimicry or emotional contagion. The top layers represent higher level processes such as perspective-taking. Nonetheless, contemporary theories argue that, if empathy was an automatic process resulting from direct perception as suggested by the PAM, then humans would be constantly empathizing \cite{heyes2018empathy,de2006empathic}. However, there are situations in which empathy is inhibited. This inhibition is produced by an adaptation resulting from a cognitive process known as \textit{empathic regulation}. For instance,  professionals in psychiatry are able to distance themselves from the patient's emotions \cite{hein2008feel}. Factors that influence empathy regulation have been widely discussed in academia \cite{wondra2015appraisal,preston2002empathy,de2007russian,de2006empathic}. In general, these regulation factors can be grouped into three categories: \textit{i)} factors related to the \textit{internal characteristics} of the \textit{observer} such as personality, mood, age, life experiences, or gender \cite{gleichgerrcht2013empathy}; \textit{ii)} \textit{situative context} factors such as the existence of more than one \textit{target} suffering different experiences, which makes it difficult to empathize \cite{de2006empathic};  and \textit{iii)} factors concerning the \textit{level of relationship} between the \textit{observer} and the \textit{target} \cite{de2006empathic}. This last category encompasses different interpersonal factors such as: \textit{similarity}, which are the perceived similarities between the \textit{observer} and the \textit{target} (e.g., gender, personality, mood, or age); \textit{familiarity}, which is related to the \textit{observer}'s previous experiences with the \textit{target} in similar situations; and the level of relationship \cite{de2006empathic} (or social tie \cite{attanasi2016social}) between the \textit{observer} and the \textit{target} derived from the interactions they have over time \cite{decety2004functional} (henceforth referred to as \textit{affective link} in order to facilitate the reading of this paper).

New advances address the phenomenon of empathy from an appraisal theory perspective \cite{heyes2018empathy,wondra2015appraisal,davis2018empathy,wellman2018theory,clark2019feel}. Under this perspective, empathy is related to higher-level cognitive processes such as other-oriented perspective-taking, self-oriented perspective-taking, and Theory of Mind \cite{wellman2018theory} (which holds that people are able to understand the mental states of others thanks to a system of rules based on their own experiences \cite{hein2008feel,gallagher2003functional}). These cognitive processes require the observer to have knowledge of affective and non-affective information about the target (e.g., beliefs, desires, and goals) \cite{goldman2011two}. In addition, empathy does not always requires the \textit{observer} to perceive the \textit{target}'s affective states but is described as a process of higher-level understanding of the \textit{target}'s situation that allows the emergence of other affective experiences (e.g., when we rejoice because someone has achieved a goal) \cite{packard2021we,argott2017acquisition}.

\subsection{Empathy in software agents}\label{empathicAgents}
The relationship of empathy with social behavior and interpersonal relationships has aroused the interest of many researchers who focuses on the areas of human behavior simulation and human-machine interaction \cite{ochs2012formal,leite2014empathic,smith2022lies}. Systems that are capable of simulating affective and empathic abilities have been used in different contexts and have proved to be more reliable and more credible
\cite{adam2009logical,paiva2017empathy,dastani2010agents}. Most of these proposals are focused on the agent-oriented programming (OAP) paradigm to develop agents with empathic abilities. For example, in the study conducted in \cite{obaid2018endowing}, a robot with empathic abilities was used in an educational environment. The study determined that the empathic robot was able to elicit and maintain the social engagement of the participants of the experiment. Similarly, the experiments conducted with empathic agents in \cite{ochs2012formal} showed that participants had a better perception of the agent when it displayed empathic emotions. Also, the research performed in \cite{rodrigues2015process} showed that agents with empathic abilities increased the involvement and sociability of participants. Therefore, considering these results, it is not surprising that, in the search for improvement in the simulation of human behavior and human-machine interaction, different proposals for computer systems with empathic abilities have appeared over the years.
For example, {\"O}.N. Yal{\c{c}}{\i}n and S. DiPaola \cite{yalcin2018computational} introduced a model of an empathic agent, based on \textit{The Russian Doll} model, composed of three layers. In the bottom layer, the \textit{communication competence} contains the processes related to the recognition and the expression of emotions. The middle layer represents \textit{emotion regulation} processes. Finally, the top layer, called \textit{cognitive mechanisms}, includes Theory of Mind and appraisal.

Another interesting approach is proposed in \cite{rasool2015empathic}. Authors present a model for an empathic agent that was based on the perception of the \textit{target}'s emotion to simulate an empathic interaction. The system has a perception layer that recognizes human emotional facial expressions. Then, the system makes an estimation of the empathic emotion by means of a regulation process that uses the agent's personality and mood as regulation factors. They use the mapping proposed in \cite{mehrabian1996analysis} to establish the relationship of personality and mood in a two-dimensional space that is based on \textit{Pleasure} and \textit{Arousal}. 

Other models focus more on the simulation of empathy through the simulation of high-level cognitive processes. For instance, the model presented in \cite{rodrigues2015process}, which also relies on the theory of De Vignemont and Singer to define the empathic process \cite{de2006empathic}, focuses on the self-projection appraisal based on OCC to elicit the empathic emotion. The appraisal process is based on a set of predefined rules that define the relationship of events to desirability. The elicited emotion is adapted by using the affective link, similarity, mood, and personality in an \textit{empathic regulation} process. Similarity is obtained by comparing the intensity and valence of both the perceived and the elicited emotions. Personality is composed of a set of thresholds for each emotion. However, these factors only affect the intensity of the emotion elicited in the empathic appraisal. Another example of the simulation of cognitive processes can be found in \cite{ochs2012formal} in which a model of an embodied empathic dialogue agent is presented. The purpose of this agent was to simulate an empathic interaction with human users. The agent deduced the user's emotions through the dialogue using a perspective-taking strategy and then responded by showing the same emotion. The agent was able to regulate the intensity of the empathic emotion based on a preset degree of empathy between the agent and the user. Experiments conducted with the agent showed that participants had a better perception of the agent when it displayed empathic emotions.

Different authors have emphasized the need to consider the inclusion of long-term interactions to improve the simulation of empathic interactions \cite{paiva2017empathy,leite2013social}. However, not much work has been done on this topic. For instance, in \cite{leite2014empathic} a model of empathy based on emotion recognition is proposed. The system is based on a robot called iCat that interacts with children while playing chess. The iCat maintains the actions that children perform in chess moves in its memory and uses this information to determine the next move taking into account the children's emotions. 

From what has been explained above, it can be deduced that most of the models of agents with empathic abilities use three fundamental processes: \textit{i)} a \textit{perception process}, in which an emotion or situation is perceived; \textit{ii)} an \textit{empathic appraisal} process, in which an empathic emotion is elicited;   and \textit{iii)} an \textit{empathic regulation} process, in which the empathic emotion is adapted to different regulation factors. Unfortunately, most of the works described above propose ad hoc methods
to simulate a certain level of affective and empathic abilities, making the agent specification highly dependent on the programmer and difficult to generalize to other domains. The use of a commonly-used agent-oriented programming language can help to improve the development of empathic agents. In addition, these languages generally include support for processes such as perception, multi-agent communication, and rational behavior. One of the most well-known agent-oriented programming languages is AgentSpeak \cite{rao1996agentspeak}. AgentSpeak is a language that is based on the logic programming paradigm for developing agents using a BDI  (beliefs, desires, and intentions) architecture. In a BDI architecture, agents are defined as intentional systems where an agent is provided with a set of mental attitudes. BDI agents are composed of a set of beliefs, desires and intentions. Beliefs represent the information that the agent has about the state of its environment. Desires are goals that the agent wants to achieve. Intentions are commitments made by the agent, i.e. those goals that the agent selects to achieve. The BDI architecture provides the basis for the development of agents based on practical reasoning. Practical reasoning is an inference process through which agents evaluate and weigh their options taking into consideration the context of the practical situation in which they are involved and their knowledge about the environment. The result of this inference process is the modification, deletion, or addition of beliefs and intentions, altering the mental state of the agent. 
There are some approaches that have used the BDI model to design agents with affective capacities \cite{boukricha2013computational,alfonso2017toward}. For example, in \cite{gluz2017probabilistic} an appraisal process based on the OCC model for eliciting emotions in BDI agents is proposed. Another interesting proposal is presented in \cite{adam2009logical}, in which a logical formalization for emotion elicitation in BDI agents based on OCC theory is presented. In the same way, a formal extension of the KARO framework \cite{meyer2006reasoning} is proposed in \cite{steunebrink2007logic} to elicit emotions in BDI agents following the OCC theory. Regarding empathy and social abilities, FAtiMA, an architecture for the development of affective BDI agents, is proposed in \cite{mascarenhas2021fatima}. This architecture uses an appraisal model based on the OCC theory for the development of empathic agents. Another interesting approach can be seen in \cite{boukricha2013computational} in which an embodied virtual agent model with empathic abilities known as EMMA (Empathic MultiModal Agent) is presented. In that model, an agent is essentially composed of a reasoning module and an empathic module. The reasoning module is based on the BDI architecture \cite{rao1995bdi}. The empathic module has a facial expression recognition system and an empathetic appraisal process that is based on the De Vignemont and Singer theory \cite{de2006empathic}. To estimate the empathic emotion, the agent attempts to internally mimic the recognized facial expression using a set of patterns called Action Units (UAs). Subsequently, the emotion is regulated in an emotion regulation process that is based on the \textit{Pleasure-Arousal-Dominance} (PAD) model \cite{mehrabian1996pleasure}. This process modifies the PAD components of emotions by means of a set of regulation factors that include similarity, mood, and deservedness (which represents the degree to which the \textit{target} deserves or does not deserve the event). 

Although there are several proposals extending AgentSpeak with different features \cite{vieira2007formal,bordini2007java,bordini2007programming}, there are very few proposals that use it when defining affective or empathic agents. For instance, in \cite{kampik2018empathic}, an empathic agent model based on AgentSpeak is presented for the resolution of conflicts of interest in interactions between agents. However, this approach focuses on utility-based functions to find solutions that are mutually acceptable, but no affective characteristic of the agents is considered in the utility functions. One of the most significant efforts to allow the use of affective agents in AgentSpeak is the {\genia} architecture \cite{alfonso2017toward}. {\genia} is a general-purpose architecture that extends AgentSpeak providing a platform that is easily adaptable to different theories of emotion in order to facilitate the development of affective agents. {\genia} adds new processes for the management of affective states, which are described in detail in Section~\ref{section:genia}. However, before the work presented here, this architecture did not provide the necessary support to the development of agents with empathic abilities.

\section{The \genia architecture}\label{section:genia}

The {\genia} architecture implements an extension of the syntax and operational semantics of the agent-oriented language AgentSpeak \cite{rao1996agentspeak}. This architecture is based on a modular design that can be easily adapted both to different psychological and neuroscience theories and to different domain-specific affective characteristics (e.g. personality or mood). To this end, {\genia} uses general components (described below) to integrate the affective processes and characteristics with in the cognitive process of a BDI agent. {\genia} offers a flexible approach that allows affective processes to influence the agent's behavior by, for example, modifying goal priorities and/or generating new intentions. In \genia, an agent is defined with its own personality, mood, and concerns (i.e., interests, motivations, ideals, or standards). This factors will influence the agent's behavior by prioritizing its goals or generating new goals. The agent's concerns together with the desirability of an event are used for the agent to decide whether the event is affective or not. Additionally, the agent also categorizes events and plans as affective using the annotation \texttt{affective\_relevant} (see Section~\ref{annotations} for the definition of annotations in AgentSpeak). Plans can also be defined using the agent's personality profile or the affective state (or affective states) as preconditions for triggering the plans. One of the main advantages of this architecture is the ability to prioritize the affective or the BDI reasoning cycles. In contrast to other affective agent architectures such as FAtiMA, {\genia} facilitates the establishment of a balance between the affective and the rational reasoning cycles. One of the personality parameters of a {\genia} agent is the rationality level, which can be specified by the programmer or estimated automatically from the values of the personality. This parameter allows to establish the priority ratio between the agent's practical reasoning processes and the affective processes. {\genia} also allows to specify how, the changes that occur in the agent's affective state, generate behaviors through an explicit and integrated coping model (see \cite{alfonso2017toward} for more details about the coping process of \genia).

The reasoning BDI cycle of {\genia} works as follows: first, in the \textit{process message} (\texttt{ProcMsg}) step, all of the pending messages received by the agent are processed; then, the \textit{select event} (\texttt{SelEv}) step selects one event to be processed taking into account all of the perceptions, messages, and stacked events. Next, the \textit{relevant plans} (\texttt{RelPl}) step retrieves all of the relevant plans considering the selected event; these plans are used to generate a list of applicable plans in the \textit{applicable plans} (\texttt{ApplPl}) step. From this list, one applicable plan is selected in the  \textit{select applicable plan} (\texttt{SelAppl}) step. Then, in the \textit{add intended means} (\texttt{AddIM}) step, all of the intended means are added to the set of intentions; the \textit{select intention} (\texttt{SelInt}) step selects one intention to be executed from the set. The selected intention is executed in the \textit{execute intention} (\texttt{ExcInt}) step. Finally, the executed intention is cleared from the set of intentions in the  \textit{clear intention} (\texttt{ClrInt}) step.

{\genia} introduced a new affective cycle that consist of five steps (dark gray shaded boxes in Figure~\ref{extensionScheme}) the \textit{event appraisal} (\texttt{Appr}) step in which the selected event is appraised; the \textit{affect adaptation} (\texttt{AffAd}) step in which the agent's mood is modified to adapt the mood according to the result of the appraisal emotion; the \textit{select coping strategies} (\texttt{Selcs}) step in which the coping strategies are selected according to the agent's mood; and the \textit{cope} (\texttt{Cope}) step, in which the coping strategies are executed. In addition, {\genia} incorporates an \textit{affective state decay process} (\texttt{AsDecay}) that continuously shifts the mood toward an equilibrium state to simulate real mood change in humans. The rate at which the mood moves to the equilibrium state can be set by the user. Finally, {\genia} modified the AgentSpeak's rational cycle to add two new steps: the \textit{affective modulation of beliefs} (\texttt{AffModB}) step in which beliefs are modified according to the agent's mood; and the \textit{evaluate expectations} (\texttt{EvalExp}) step, which includes the possibility to add temporal expectations to agents \cite{taverner2016integrating}. {\genia} allows to modify the agent's behavior to simulate affective behaviors. For this purpose, {\genia} modifies the selection of applicable plans, considering the agent's \textit{rationality level}. {\genia} uses two separate sets of applicable plans: a set of applicable plans $R$ selected by the BDI practical reasoning cycle, which considers the beliefs and desires and intentions of the agent; and a set of applicable plans $A$ that, in addition to beliefs, desires, and intentions, also consider, agent's concerns and affective characteristics such as personality, emotion, or mood. Subsequently, the final applicable plan is selected, in the  \textit{select applicable plan} (\texttt{SelAppl}) step, using the equation:

\begin{gather}
    \texttt{applicable\_plan} = \underset{p\in (R~\cup~A)}{\arg\max}\,\left\{\begin{matrix*}[l]
\textit{Priority}(p) \cdot \textit{rl}, & \text{if}~p\in R \\
\textit{Priority}(p) \cdot (1-\textit{rl}~), & \text{if}~p\notin R \\
\end{matrix*}\right.
\end{gather}

that considers the priorities $\textit{Priority}(p)$ of the plan $p\in (R~\cup~A)$ and the rationality level \textit{rl} of the agent.

To facilitate the development of affective agents, {\genia} incorporates a default design that can be adapted by advanced users interested in customizing the behavior of the agent. In this default design, the user can choose between two different appraisal models: a model based on OCC theory \cite{ortony1990cognitive} that uses the appraisal variables defined in \cite{marsella2009ema} (i.e., desirability, likelihood, and causal attribution) using a numerical approach; and a model based on Scherer's theory using fuzzy logic to define the appraisal rules \cite{tavernerIS}. {\genia} also offers the possibility of selecting between two models of representation of affective states: a model based on the PAD \cite{mehrabian1996pleasure} described in \cite{alfonso2017toward}; and a multidimensional culturally adapted representation of emotions presented in \cite{taverner2020multidimensional}. Finally, in this default design, the personality is defined by the OCEAN model \cite{mccrae1992introduction}.

\section{{\egenia} an AgentSpeak extension for empathic agents}\label{proposal}
In this section, we present {\egenia} an extension of the  syntax, semantics, and the reasoning cycle of an AgentSpeak agent to support the development of agents with empathic abilities. This extension  allows emotions and empathy to have an effect on the practical reasoning of the agent. Emotions and empathy will be involved in an affective cycle that will affect that will affect the agent's reasoning processes. To do that, we have extended the {\genia} architecture with a new syntax and semantics to facilitate the emergence of empathy in software agents. {\egenia} includes the three fundamental processes identified in the literature on empathic agents discussed in Section~\ref{empathicAgents}: perception, empathic appraisal, and empathic regulation (see Figure~\ref{extensionScheme}). The perception process is an intrinsic part of AgentSpeak. However, we have added a new process to differentiate empathic affective events from other affective events, the \textit{event classification} (\texttt{EvClass}) process. This process consists of two phases. In the first phase, an event is perceived following the process described in \cite{bordini2007programming}. In the second phase, the event is evaluated to determine whether or not it is an affective event. This second phase defines the agent's cognitive ability to maintain a sense of self as distinct from the \textit{target} agent to elicit a target-oriented empathic emotion. This cognitive ability is simulated using two different appraisal processes: one for empathic events (\texttt{EmphAppr}) and the other for non-empathic affective events (\texttt{Appr}). 
On the other hand, empathic appraisal is composed of a self-projection appraisal, which evaluates the event using the agent's own beliefs, concerns, and affective memory. The \textit{empathic regulation}  process (\texttt{EmphReg}) adapts the empathic emotion to the agent's affective characteristics (e.g., mood or personality) and the knowledge that the agent has about the \textit{target} (e.g., affective link or trust level). Finally, the \textit{emotion selection} process (\texttt{EmSel}) selects the most appropriate emotion. 


\subsection{A simple example}\label{simpleexample}
\begin{figure}[bt]
\centering
\includegraphics[trim = 245mm 167mm 240mm 167mm, clip, width=90mm]{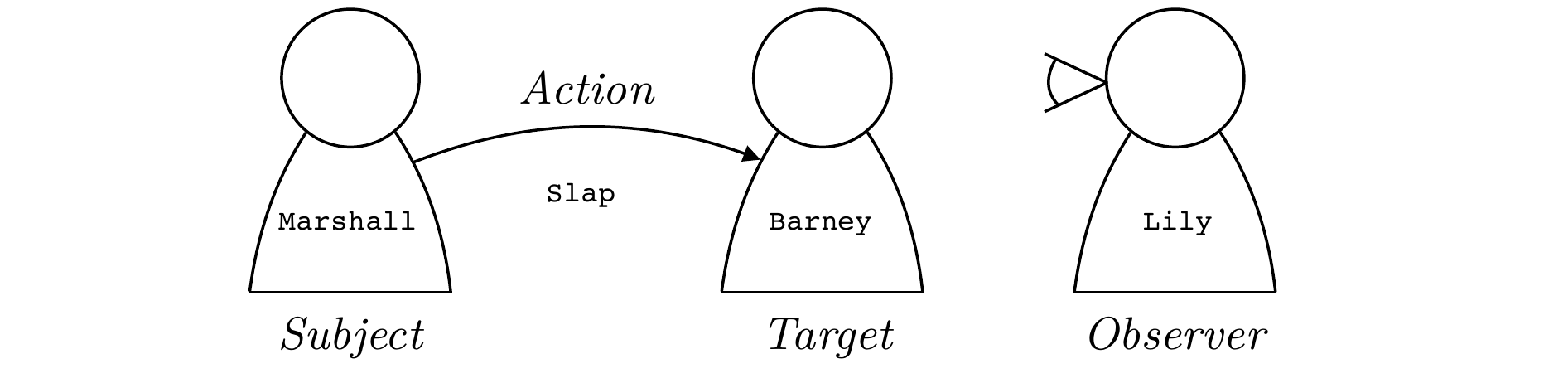}
\caption{Participants involved in an empathic event.}
\label{participants}
\end{figure}

From the theories of empathy previously cited, it can be assumed that, for empathy to occur, there must be at least two actors: one who suffers in a situation (the \textit{target}) and another who perceives the situation and reacts empathetically (the \textit{observer} or empathic agent). To make the reading more convenient, we illustrate the problem with an example. Let us assume a more general scenario with three actors: Marshall, Lily, and Barney. At one point in time Marshall slaps Barney in the face. 
This event causes Barney to become sad. Lily, who is in the same room, sees the entire scene and empathizes with Barney and feels sorry for what has happened to him. 

In this scenario, there are basically two interactions: first, Marshall performs an action directed to Barney; second, Lily perceives what has happened to Barney. Three different roles of agents can be identified in these interactions. Marshall, the agent that performs the action, henceforth known as the \textit{subject} agent; Barney, the agent that receives the consequences of the action, henceforth known as the \textit{target} agent; and Lily, who perceives the action, henceforth known as the \textit{observer} agent, and that can potentially empathize with the \textit{target} of the event. 

\subsection{Formalization of an empathic agent}\label{sec:configuration}

\begin{figure}
    \centering
    \includegraphics[trim = 14mm 90mm 12mm 90mm, clip, width=\columnwidth]{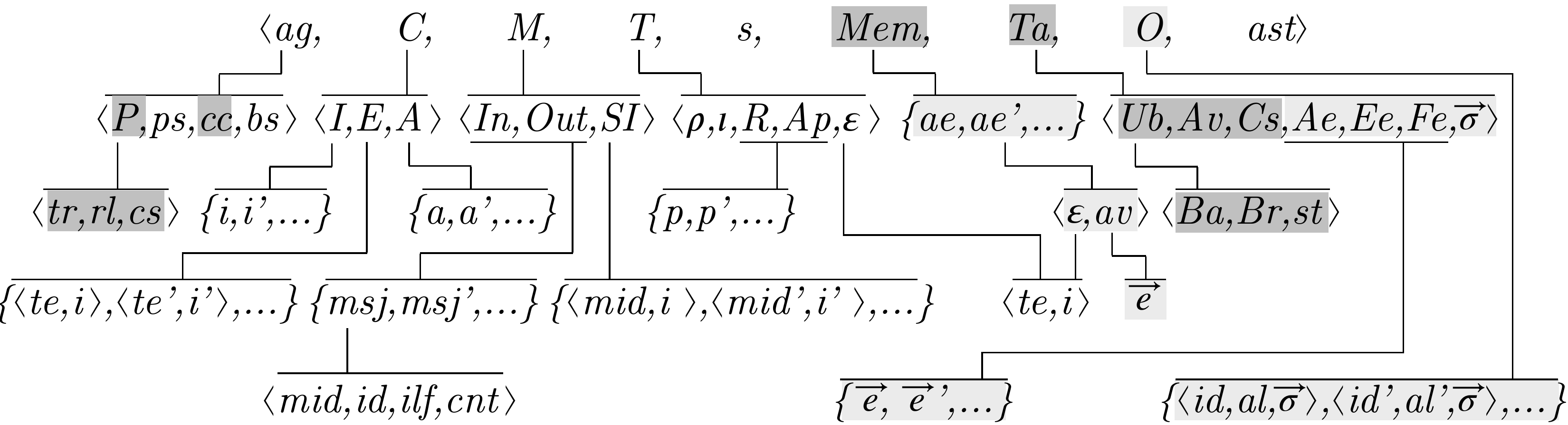}
    \caption{{\egenia} agent configuration.}
    \label{newConfiguration}
\end{figure}

In \cite{rao1996agentspeak}, the agent-oriented language AgentSpeak semantics is defined using an operational semantics formalism. This operational semantics defines the structure and the configuration of the agent program and the transitions derived from its internal reasoning. In this section, we extend this operational semantics to define the new {\egenia} configuration for empathic agents. 

Figure~\ref{newConfiguration} shows the new configuration of a {\egenia} agent. The dark gray shaded attributes are the ones that {\genia} added to the original configuration proposed for AgentSpeak in \cite{bordini2007java}. The light gray shaded attributes represent the new components of {\egenia} to add information that is relevant to the development of empathic agents. In this new configuration, an agent is defined by a tuple $\left \langle \textit{ag},C,M,T,s,\textit{Mem},\textit{Ta},\textit{O},\textit{ast} \right \rangle$, where:
\begin{itemize}
\renewcommand\labelitemi{--}
    \item \textit{ag} is the set of attributes that constitute the agent defined by the tuple $\left \langle P,\textit{ps},\textit{cc},\textit{bs} \right \rangle$, where:
    \begin{itemize}
        \item[--] \textit{P} is the agent's personality represented by the tuple  $\left \langle \textit{tr},\textit{rl},\textit{cs} \right \rangle$, where: 
        \begin{itemize}
            \item[--] \textit{tr} is a set of personality traits. Each personality trait is a value representing the level that the agent has of that personality trait. For example, when using the OCEAN model, five personality traits are defined, one for each component of the OCEAN model.
            
            \item[--] \textit{rl} is a value representing the agent's rationality level. The higher the rationality level, the higher the priority of plans activated in a rational BDI cycle, making the agent more rational. However, the lower the rationality level, the higher the priority of plans activated by the affective cycle, making the agent more emotional.
            
            \item[--] \textit{cs} is the agent's set of coping strategies. These coping strategies relate affective states and beliefs to a set of intentions that will be included in the agent's agenda.
            
        \end{itemize}
        \item[--] \textit{ps} and \textit{bs} represent the set of plans and the set of beliefs of the agent, respectively.
        
        \item[--] \textit{cc} is the set of concerns of the agent that represents the motivations, standards, ideals and/or interests of the agent.
        
    \end{itemize}
    \item \textit{C} is the current circumstance represented by a tuple composed of: $I$, which is the set of intentions $\{i,i',\cdots\}$; $E$, which is a set of events composed of a set of tuples $\left \langle \text{triggering event } \textit{te}, \text{ intention } i \right \rangle$; and $A$, which is a set of actions $\{a,a',\cdots\}$. 
    
    \item \textit{M} are the communication parameters represented by the tuple $\left \langle \textit{In},\textit{Out},\textit{SI} \right \rangle$, where \textit{In} and \textit{Out} represent the list of input and output messages, respectively and \textit{SI} is a set of suspended intentions composed by a set of tuples $\left \langle \text{message identifier } \textit{mid}, \text{ intention } i \right \rangle$. Each message \textit{msj} is composed of the message identifier \textit{mid}, the identifier of the agent which sent the message \textit{id} (i.e., the \textit{sender} agent), the illocutionary force \textit{ilf}, and the message content \textit{cnt}.
    
    \item \textit{T} is the temporary information of the current rational cycle consisting of a tuple containing: an applicable plan $\rho$, a particular intention $\iota$, the sets of relevant plans $R$, the set of applicable \textit{Ap} plans composed of a set of plans $\{p,p',\cdots\}$, and the event $\varepsilon$ that triggered the rational cycle  represented by a tuple $\left \langle \text{triggering event } \textit{te}, \text{ intention } i \right \rangle$. 
    \item \textit{s} is the current step of the rational cycle where:
    \begin{gather}
    \begin{aligned}
    \textit{s} \in & \left\{ \texttt{ProcMsg}, \texttt{AffModB}, \texttt{EvalExp}, \texttt{SelEv}, \texttt{RelPl},\right.\\
    &\left.\texttt{ApplPl}, \texttt{SelAppPl},\texttt{AddIm}, \texttt{SelInt}, \texttt{ExcInt}, \texttt{CrlInt}\right\}
    \end{aligned}
    \end{gather}
    
    \item \textit{Mem} is the affective memory that, in the original {\genia} architecture, consists of a set of events. In {\egenia}, %
    it consists of a set of affective events \textit{ae}. An affective event \textit{ae} is defined as a tuple $\left \langle \text{event } \varepsilon, \text{ affective value } \textit{av} \right \rangle$, where the affective value \textit{av} is an attribute that represents the emotion that the event $\varepsilon$ produced in the agent.
    
    \item \textit{Ta} represents the temporal information of the affective cycle and is composed of: \textit{Ub}, which is a tuple containing the set of beliefs that are going to be added to the belief base \textit{Ba}, the set of beliefs that are going to be removed from the belief base \textit{Br}, and the identifier of the step \textit{st} of the cycle in which the beliefs are going to be added or removed; \textit{Av}, which is the set of appraisal variables; \textit{Cs}, which is the set of coping strategies to be executed; \textit{Ae}, which is the set of emotions that can be elicited by the \textit{appraisal} process; \textit{Ee}, which is the set of empathic emotions that can be triggered by the \textit{empathic appraisal} process; \textit{Fe}, the final emotion (or emotions) resulting from the \textit{emotion selection} process; and $\vec{\sigma}$, which represents the current mood of the agent. The emotions contained in \textit{Fe} will be the ones considered as active in the affective cycle. Emotions and mood are represented by a n-dimensional vector $\vec{e}$. For example, for the PAD model, this vector will have three components (i.e. one for each dimension). 
    
    \item \textit{O} is a new component which is added to represent the information that the empathic agent knows about other agents in the environment. This information is composed of a set of tuples each of which corresponds to one agent in the environment. Each tuple contains the agent identifier \textit{id}, the affective link \textit{al} that the empathic agent has with the agent identified by \textit{id}, and the \textit{id} agent's mood $\vec{\sigma}$. This tuple can be extended in the future to contain more knowledge about agents in the environment, such as their concerns, goals, beliefs, or trust level. The affective link is a value that indicates the affective link between both agents. The greater the affective link value, the greater the relationship between the agents. Negative affective links indicate enmity between the agents. The affective link can be modified due to the interactions between the agents. 
    \item \textit{ast} is the current step of the affective cycle, where:
    \begin{gather}
    \begin{aligned}
    \textit{ast} \in & \left\{ \texttt{EvClass}, \texttt{Appr}, \texttt{EmphAppr}, \texttt{EmReg}, \right.\\
    &\left. \texttt{EmphReg}, \texttt{EmSel}, \texttt{AffAd}, \texttt{SelCs}, \texttt{Cope}\right\}
    \end{aligned}
    \end{gather}
\end{itemize} 

\subsection{Extending the AgentSpeak language to identify the actors of an empathic interaction}\label{annotations}

In AgentSpeak, an agent is defined as a set of beliefs and a set of plans. The knowledge contained in the belief base may not necessarily be complete or accurate, since the environment may be very large and may experience changes that the agent has not perceived. On the other hand, plans contain basic actions that the agent can perform to change its environment. Plans are composed of a triggering event, a context, and a set of sequential instructions that may include: updates to the belief base, actions, or goals. Triggering events refer to the addition or deletion of beliefs or goals. In AgentSpeak, both beliefs and goals are defined as atomic formulas \cite{rao1996agentspeak}. An atomic formula representing a belief or a goal is composed of a predicate and a set (possibly empty) of $n$ terms of a first order logic: 

\begin{gather}
\texttt{predicate}(\texttt{term}_1,\texttt{term}_2,\cdots,\texttt{term}_n)
\end{gather}

For example, the triggering event \texttt{time(cloudy)} is composed of the predicate \texttt{`time'} and the term \texttt{`cloudy'}.  

In \cite{vieira2007formal} and \cite{bordini2007java}, the concept of annotation was added to AgentSpeak to provide the ability to express properties associates with events and beliefs. The syntax for an atomic formula with annotation in AgentSpeak is: 

\begin{gather}
\texttt{predicate}\left(\texttt{term}_1,\texttt{term}_2,\cdots,\texttt{term}_n\right)\left[\texttt{a}_1,\texttt{a}_2,\cdots,\texttt{a}_m\right]
\end{gather}

where each $a_i$ represents the $i$th annotation defined as:

\begin{gather}
\texttt{a}_i = \texttt{functor}_i\left(\texttt{term}'_{i,1},\texttt{term}'_{i,2},\cdots,\texttt{term}'_{i,n}\right)
\end{gather}

where an atom (called functor) is followed by a number of terms (called arguments). $\texttt{term}'_{i,j}$ is the $j$th term of the annotation $\texttt{a}_i$. 

This extension of the language provides more expressiveness to AgentSpeak allowing different types of properties to be defined. For example, we can add one annotation to an event to represent the sources of the event. The keyword that we use to represent this annotation is "source". For example, in the triggering event:

\begin{gather*}
\texttt{time}\left(\texttt{cloudy}\right)\left[\texttt{source}(\texttt{Marshall})\right]
\end{gather*}

the annotation \texttt{source}(\texttt{Marshall}) indicates that the source of the triggering event is the agent Marshall.

In general, there are three possible sources for an event: 
\begin{itemize}
\renewcommand\labelitemi{--}
    \item \textit{perceptions}, which represent the information that the agent perceives from its environment. Perceptions are represented by the annotation \texttt{source(percept)}. 
    \item \textit{mental notes}, which represent beliefs that the agent acquires or deduces by itself, such as memories or changes in the agent's state. The mental notes are expressed through the annotation \texttt{source(self)}.
    \item \textit{communications}, which is the information that comes from another agent of the system as a consequence of a communication act.   
\end{itemize}

In our proposal, we have extended AgentSpeak with new annotations that allow the events with relevant information for empathy to be contextualized for its simulation. As we discussed in Section \ref{relatedWork}, the \textit{empathic regulation} process is affected by a set factors that include interpersonal factors related to the \textit{target} agent such as the affective link, concerns, goals, similarity, or trust level. Therefore, to elicit an empathic emotion, it is necessary to identify the \textit{target} agent when an event occurs. This information can be implicit in the semantics of the event, in which case the agent may deduce the \textit{target} of the action through an inference process, or it can be explicitly incorporated into the syntax of the triggering event. We have used this second approach because it simplifies the agent's programming by easily identifying the \textit{subject} and the \textit{target} of an event.
We have extended the representation of a triggering event to explicitly include the agents involved in an action (i.e., \textit{subject} and \textit{target} agents) without modifying the original syntax of the atomic formula of AgentSpeak. This has been achieved using two annotations to identify the agents participating in the action represented by the triggering event: the \texttt{subject} annotation, which identifies the \textit{subject} agent that performs the action; and the \texttt{target} annotation, which identifies the \textit{target} agent receiving the consequences of the action. Following this definition, a triggering event will be represented by the structure:

\begin{gather}
\begin{aligned}
&\texttt{predicate}\left(\texttt{term}_1,\texttt{term}_2,\cdots,\texttt{term}_n\right)\\
&\left[ \texttt{subject}(\textit{subject\_id}),\texttt{target}(\textit{target\_id}),\texttt{a}_3,\texttt{a}_4,\cdots,\texttt{a}_n\right]
\end{aligned}
\end{gather}

where \textit{subject\_id} and \textit{target\_id} are the identifiers of the \textit{subject} agent and the \textit{target} agent, respectively. Note that, the order in which the annotations are defined is not relevant since the annotations can be written in any order. By adding these new annotations, it is now possible to identify the agents involved in any action. Based on the example of Section~\ref{simpleexample}, we can use the following expression to specify the triggering event ``Marshall has slapped Barney'':

\begin{gather*}
\texttt{slap[subject(Marshall),target(Barney)]}
\end{gather*}

This event has a predicate ``slap'' and two annotations: \texttt{subject(Marshall)} and  \texttt{target(Barney)}. Moreover, we propose adding an optional annotation to include the value of the interaction. The interaction value of a triggering event is an optional number that is associated with the triggering event in the range of $\left[-1,1\right]$, which identifies if the interaction is good (positive interaction value), bad (negative interaction value), or neutral (interaction value equals $0$) for the agent that receives the action. The interaction value is used to update the affective link between the agents. A positive interaction value indicates that the interaction has a positive effect on the affective state of the \textit{target} agent, improving the affective link that the \textit{target} agent has with the \textit{subject} agent. A negative interaction value has a negative impact in the agent's affective state and decreases the affective link. An interaction value equal to $0$ denotes that the interaction has a neutral effect on the affective state of the agent. Therefore, it has no effect on the affective link. Note that, if necessary, the interaction value can also be deduced automatically by the agent through the appraised emotion (for instance, considering the level of pleasure of the generated emotions) or through an internal inference process.


We have introduced new functions to handle the annotations that will be used in Section~\ref{formalization} to formalize the new internal processes of the affective cycle of an agent. First, the function to obtain the subject agent of a triggering event \textit{te} is defined as: 

\begin{gather}
    \texttt{getSubject}(\textit{te}) := \left\{\begin{matrix*}[l]
    \textit{term}_{i}, & \text{if}~\exists~a_i\in\textit{te}_{annots}:\textit{functor}_i =\\
    &\texttt{`subject'} \text{and}~\textit{term}_{i}\neq \texttt{ag\_id} \\
    \texttt{self}, & \text{if}~\exists~a_i\in\textit{te}_{annots}:\textit{functor}_i =\\
    &\texttt{`subject'}\text{and}~\textit{term}_{i} = \texttt{ag\_id}\\ 
    \texttt{null}, & \text{otherwise} 
\end{matrix*}\right.
\label{eq:getSubject}
\end{gather}

where $\textit{te}_{annots}$ is the set of annotations associated to the triggering event $\textit{te}$, $a_i$ is the $i$th annotation of the set of annotations $\textit{te}_{annots}$, \textit{functor} is the functor of the annotation $a_i$, $\textit{term}_i$ is the term of the annotation $a_i$, \texttt{`subject'} is a keyword that identifies the \textit{subject} annotation, and \texttt{ag\_id} is the agent's identifier. If $\textit{term}_i$ is equal to the \texttt{ag\_id}, it indicates that the triggering event comes from the agent itself. If $\textit{term}_i$ is not the \texttt{ag\_id}, it indicates that the triggering event comes from another agent. Finally, if there is no \texttt{`subject'} annotation, it indicates that this triggering event comes from an unknown subject.

Similarly, the \textit{target} of a triggering event \textit{te} is obtained through the function:

\begin{gather}
    \texttt{getTarget}(\textit{te}) := \left\{\begin{matrix*}[l]
    \textit{term}_{i}, & \text{if}~\exists~a_i\in \textit{te}_{\textit{annots}}:\textit{functor}_i =\\
    &\texttt{`target'}\\
    \texttt{null}, & \text{otherwise} 
\end{matrix*}\right.
\end{gather}

where \texttt{`target'} is a keyword that identifies the \textit{target} agent annotation. 

Finally, to obtain the interaction value of the triggering event \textit{te}, we define the function:

\begin{gather}
\begin{aligned}
\texttt{getIV}(\textit{te}) := \left\{\begin{matrix*}[l]
    \textit{term}_{i}, & \text{if}~\exists~a_i\in \textit{te}_{\textit{annots}}: \\
    &\textit{functor}_i = \texttt{`InteractionValue'}\\
    \texttt{0}, & \text{otherwise}
\end{matrix*}\right.
\end{aligned}
\end{gather}

where \texttt{`InteractionValue'} is a keyword that identifies the annotation that contains the interaction value. 

\subsection{Extending the AgentSpeak agent configuration}\label{formalization}\label{syntax}

\begin{figure}[bt]
\footnotesize{
\setlength{\grammarparsep}{0pt plus 1pt minus 1pt} 
\setlength{\grammarindent}{9em}  
\begin{grammar} 
<agent>   $\rightarrow$ \myttuline{init_beliefs} "["\myttuline{concerns}"]" "["\myttuline{personality}"]" "["\myttuline{others_info}"]" \myttuline{init_goals}  \myttuline{plans}

<others_info> $\rightarrow$ `others__:' `[' \myttuline{other} ( `,' \myttuline{other} )* `]'

<other> $\rightarrow$ \myttuline{ag_id} `:' `[' \myttuline{list_attr} `]'

<ag_id> $\rightarrow$ \myboldtt{\texttt{ATOM}}

<list_attr> $\rightarrow$ ( \myttuline{attr_label} `:' ( \myboldtt{\texttt{NUMBER}} |  \myboldtt{\texttt{ATOM}} ) (`,' ( \myboldtt{\texttt{NUMBER}} |  \myboldtt{\texttt{ATOM}} ) )*

<attr_label> $\rightarrow$ \myboldtt{\texttt{ATOM}}

<personality>   $\rightarrow$  `personality__:' `\{'  \myttuline{list_traits} "["`,' \myttuline{rat_level}"]" "["`,' \myttuline{coping_strats}"]" `\}'`.'

<list_traits> $\rightarrow$ `[' \myttuline{trait} ( `,' \myttuline{trait} )* `]'

<trait> $\rightarrow$ \myttuline{trait_label} `:'  \myboldtt{\texttt{NUMBER}}

<trait_label> $\rightarrow$ \myboldtt{\texttt{ATOM}}

\end{grammar}}
\caption{Simplified extension of the agent's syntax including the new extension of \egenia.}
\label{agentEBNF}
\end{figure}

\begin{figure}
\footnotesize{
\setlength{\grammarparsep}{0pt plus 1pt minus 1pt} 
\setlength{\grammarindent}{9.5em}  
\begin{grammar} 
<mas> $\rightarrow$ `MAS' \myboldtt{\texttt{ID}} `\{' "["\myttuline{infrastructure}"]" "["\myttuline{environment}"]" \\
"["\myttuline{execcontrol}"]" "["\myttuline{w_matrix}"]" `\}'

<w_matrix> $\rightarrow$ `w_matrix__:' \myttuline{w_weights} ( `,' \myttuline{w_weights} )*

<w_weights> $\rightarrow$ `[' \myttuline{trait_label} `:'  \myttuline{list_weights} `]' 

<trait_label> $\rightarrow$ \myboldtt{\texttt{ATOM}}

<list_weights> $\rightarrow$ `[' \myttuline{weight} ( `,' \myttuline{weight} )* `]'

<weight> $\rightarrow$ \myttuline{em_label} `:' \myboldtt{\texttt{NUMBER}}

<em_label> $\rightarrow$ \myboldtt{\texttt{ATOM}}
\end{grammar}}
\caption{Simplified extension of the MAS project syntax including the new extension of \egenia.}
\label{masEBNF}
\end{figure}

The syntax of AgentSpeak was presented in \cite{bordini2007programming} using the EBNF (Extended Backus-Naur Form) notation \cite{backus1963revised}. This syntax was later extended by the {\genia} architecture to allow the development of affective agents  \cite{alfonso2017toward,taverner2018modeling,taverner2016integrating}. In the original EBNF syntax of {\genia}, an agent (\texttt{agent}) is defined by the set of initial beliefs (\texttt{init\_beliefs}), concerns (\texttt{concerns}), personality (\texttt{personality}), initial goals (\texttt{init\_goals}), and plans (\texttt{plans}). We have extended this syntax to incorporate some new attributes to the agent configuration. Figure~\ref{agentEBNF} and~\ref{masEBNF} shows the extension of the EBNF syntax\footnote{Definitions of non-terminal syntactic symbols that have not been described in this section are detailed in \cite{bordini2007programming} and \cite{alfonso2017toward}.}. We have added a new attribute to represent the knowledge that the agent has about other agents (\texttt{others\_knowledge}). To represent this knowledge, we use a set consisting of the agent's identity (\texttt{ag\_id}) and a list of attributes associated with that agent (\texttt{list\_attr}). The list of attributes is defined as a set of tuples consisting of the attribute label (\texttt{attr\_label}) and its value. One of the most important attributes is the affective link that represents the affective proximity or relationship between the agents. In our default design the affective link is represented as a value ranging from [-1,1], but it can be easily modified to use other values such as fuzzy values. Following the example described in Section~\ref{simpleexample}, let us assume that agent Marshall has an affective link with agent Barney of $-0.5$ and an affective link with agent Lily of $0.9$. To represent this knowledge, the following sentences must be added to agent Marshall's definition: 

\begin{verbatim}
others__: [ Lily: [ affective_link: 0.9 ], 
        Barney: [ affective_link: -0.5 ] ]
\end{verbatim}

Note that the \texttt{affective\_link} is not the only attribute that can be represented. The number of attributes could be increased to represent more information such as trust level or the other agents' goals, concerns, and beliefs. 

On the other hand, in {\genia}, the agent's personality is defined by the keyword ``\texttt{personality\_\_:}" followed by some attributes: the traits (\texttt{traits}), which are defined as a list containing a value for each personality trait; and optionally, the rationality level (\texttt{rat\_level}), which is defined as a numerical value representing how rational the agent is; and the coping strategies (\texttt{coping\_strats}), which are defined as a list of plans that will be used by the agent to deal with affective events. We have modified the definition of the personality traits in {\egenia} to make it more user friendly. Now, the personality traits are defined using a trait label (\texttt{trait\_label}) followed by the numeric value for that trait. For example, let us suppose that the agent Marshall has an \textit{Extraversion} level of $0.9$ and a \textit{Neuroticism} level of $0.1$. We can express this using the new syntax as follows:

\begin{verbatim}
personality__:  { [ extraversion: 0.9, 
    neuroticism: 0.1 ] }
\end{verbatim}

Finally, we have introduced a new attribute $\omega$ (\texttt{em\_weights}) to the multi-agent system (MAS) project. $\omega$ represents a correlation matrix between the personality traits $\textit{tr}\in P$ and all of the possible emotion types $t\in{\textit{Ae}\cup\textit{Ee}}$, where \textit{Ae} is the set of emotions that can be elicited by the \textit{appraisal} process and \textit{Ee} is the set of empathic emotions that can be triggered by the \textit{empathic appraisal} process. For each pair of values $(\textit{tr},t)$, a weight $p$ representing how the personality trait $tr$ influences the emotion type $t$ can be specified, but many of these weights will be zero. Therefore, the set of weights of the emotions can be viewed as the set of correlations between emotions and personality traits: the greater the correlation between the personality trait $tr$ and the emotion type $t$, the greater the value of $p$. For example, considering that the trait of \textit{Extraversion} $E$ is related to positive emotions and \textit{Neuroticism} $N$ does not have a very high relation with positive emotions \cite{furnes2019exploring}, for the \textit{Happiness} emotion, the \textit{Extraversion} weight ($\omega_{Happiness,E}$) will be greater than the weight of the \textit{Neuroticism} for the  \textit{Happiness} emotion ($\omega_{Happiness,N}$).

To define the syntax of the $\omega$ matrix, we use the keyword ``\texttt{w\_matrix\_\_:}" followed by one or more weights (\texttt{w\_weights}). These weights are defined as a label that identifies the personality trait (\texttt{trait\_label}) followed by a list of weights (\texttt{list\_weights}). This list of weights is composed of one or more emotion labels (\texttt{em\_label}) followed by a number that represents the correlation between that personality trait and that emotion. For instance, let us consider the following example in which a personality / emotion correlation matrix is defined based on the variables \textit{Extraversion} and \textit{Neuroticism} and the emotions \textit{Anger} and \textit{Sadness} (as proposed in \cite{tavernerIS}). Table~\ref{corMatrix} shows the correlation matrix. The values of this correlation matrix are based on the results obtained from the experiments presented in \cite{furnes2019exploring}. The definition of the $\omega$ matrix using the proposed syntax is:

\begin{verbatim}
w_matrix__: [ extraversion: [ Anger: 0.5, 
                Sadness: 0.6 ],
            neuroticism: [ Anger: 0.8, 
                Sadness: 0.7 ] ]  
\end{verbatim}

This correlation matrix is defined in the configuration file (i.e., MAS project file) of the multiagent system and is common to all agents defined in that environment. Note that, in the default design of {\genia}, the OCEAN personality model is used, but the architecture also allows the use of other personality models with numeric or fuzzy variables as presented in \cite{taverner2018modeling}. Through our extension of AgentSpeak, the programmer can clearly define both the personality of the agent and the knowledge that the agent has about other agents. On the one hand, instead of establishing one threshold for each emotion, as seen in previous approaches, our model allows the relationship between different personality traits and emotions to be defined using a correlation matrix $\omega$. Therefore, just by varying the personality traits we can define agents with different affective behaviors. On the other hand, in our proposal, each agent could have its own affective link matrix, thus allowing the implementation of asymmetric relationships between agents. This difference in the affective links between agents represents what happens in human societies in real life. In addition, the affective link between two agents can dynamically vary according to the interaction value and the emotion generated when the agents interact with each other. This allows the simulation of affective behavior in long-term interactions between agents.  


\begin{table}[bt]
\centering
\caption{Example of a personality / emotion correlation matrix.}\label{corMatrix}
\begin{tabular}{lcc}
\hline
Emotions        & Extraversion & Neuroticism \\ 
\hline
Anger   & 0.5          & 0.8         \\
Sadness & 0.6          & 0.7         \\ 
\hline
\end{tabular}
\end{table}

\subsection{
A new rational cycle for an empathic BDI agent}\label{sec:semantic}
\begin{figure*}[bt]
\centering
\includegraphics[trim = 0mm 89mm 0mm 90mm, clip, width=\textwidth]{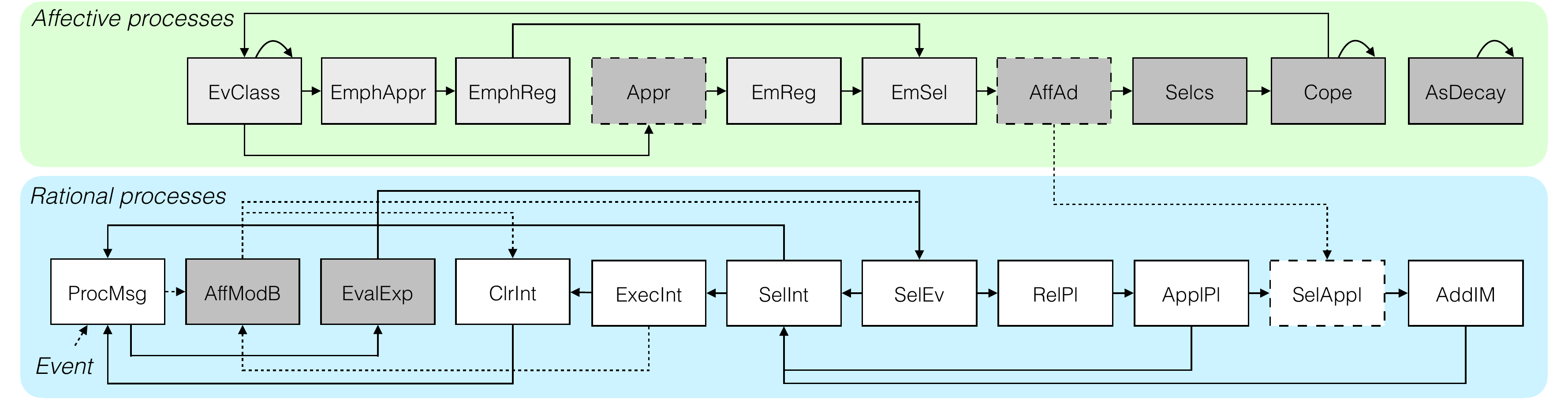}
\caption{New configuration of both affective and rational processes. White boxes represent the original processes proposed for an AgentSpeak agent; dark gray shaded boxes are the original processes of the {\genia} architecture; dashed lines indicate processes that have been redesigned in {\egenia}; light gray shaded boxes represent the new processes of {\egenia}.}
\label{extensionScheme}
\end{figure*}

We have redesigned the {\genia} architecture by adding new processes to enable the development of empathic agents. Figure~\ref{extensionScheme} shows the new configuration of {\egenia}. We have introduced five new steps to the affective cycle and we have proposed a new formalization of the transitions between these steps using the operational semantics described in \cite{plotkin1981structural}. This operational semantics consists of a set of rules that define the transitions between different configurations of an AgentSpeak agent.

To make readability more convenient, we have adopted the syntax proposed in \cite{vieira2007formal}. Following this syntax, if \textit{ag} is the set of attributes that define an agent, we will refer to the personality component $P$ of \textit{ag} as $\textit{ag}_P$. We will also refer to the empathic agent whose affective cycle we are describing as \textit{empathic} agent. In addition, we will use \textit{emph\_ag\_id} as the empathic agent identifier.

\subsubsection{Event classification }
The \textit{event classification } (\texttt{EvClass}) step evaluates the triggering event \textit{te} component of the current event $T_\varepsilon$ to determine whether the event is affectively relevant by a Boolean function of degree $n$:

\begin{gather}
\texttt{affRelEv}(\textit{te}) := X^n \to X
\end{gather}

where $X=\{\textit{True},\textit{False}\}$ is a Boolean domain. This function is defined by default as:

\begin{gather}
\texttt{affRelEv}(\textit{te}) := \left\{\begin{matrix*}[l]
\texttt{True}, &  \text{if}~\exists~a_i\in\textit{te}_\textit{annots}:\textit{functor}_i =\\
& \texttt{`affective\_relevant'}  \\ 
\texttt{False}, & \text{otherwise} 
\end{matrix*}\right.
\end{gather}

where \textit{te} is the triggering event, $\textit{te}_\textit{annots}$ is the set of annotations of \textit{te}, $a_i$ is the $i$th annotation of the set of annotations $\textit{te}_\textit{annots}$, and $\textit{functor}_i$ represents the functor of the annotation $a_i$. 

Furthermore, this step also performs the dissociation between events that should be evaluated in the affective cycle and those that should be evaluated in the empathic cycle allowing the distinction between the self from the \textit{target}. As discussed in Section~\ref{relatedWork}, this distinction between self and \textit{target} is essential in the development of empathy since it allows an agent to differentiate its own emotions from emotions produced by the perception of others. For an event, we have designed four transition rules in this step. The first rule (EvClass$_1$) is applied to events that are not affectively relevant and either have nothing to do with the \textit{empathic} agent (e.g., the \textit{target} are other agents) or events whose \textit{target} is the \textit{empathic} agent but without a \textit{subject} agent (e.g., environmental events): 

\begin{gather*}
    \frac{\neg \texttt{affRelEv}(\textit{te}) \; \wedge \; (\textit{tg} \neq   \textit{\textit{emph\_ag\_id}} \; \vee \; (\textit{tg} =   \textit{\textit{emph\_ag\_id}} \; \wedge \; \textit{sbj} \in \left\{\texttt{null},\texttt{self}\right\}))}{\left \langle \textit{ag},C,M,T,s,\textit{Mem},\textit{Ta},\textit{O},\texttt{EvClass} \right \rangle \rightarrow \left \langle \textit{ag},C,M,T,s,\textit{Mem},\textit{Ta},\textit{O},\texttt{EvClass} \right \rangle} \tag{EvClass$_1$}
\end{gather*}
\begin{align*}
    \textit{where: }
    &T_{\varepsilon} = \left \langle \textit{te},i\right \rangle, \text{$T_{\varepsilon}$ is the $\varepsilon$ component of the tuple $T$ }\\
    &\text{following the syntax presented in \cite{vieira2007formal} (see Figure~\ref{newConfiguration}).}\\
                    &\textit{tg}=\texttt{getTarget}(\textit{te})\\
                    &\textit{sbj} = \texttt{getSubject}(\textit{te})
\end{align*}

If this rule is evaluated as true, the \textit{empathic} agent will remain in the state \texttt{EvClass} waiting for a new event. For example, let us assume that Lily is the \textit{empathic} agent. At a certain point, in time Lily perceives the event \texttt{time(cloudy)}. The function \texttt{affRelEv} evaluates the event obtaining the value false as a result because this event is not relevant for the affective cycle. The function \texttt{getTarget} also evaluates the event and returns the value {\texttt{null}} since there is no annotation indicating a \textit{target} agent at the event. Therefore, the event does not have any affective impact on Lily, and the affective cycle is not started. Lily will remain in the \textit{event classification} step waiting for a new event. 

The second rule (EvClass$_2$) deals with events that are not affectively relevant but that have some level of social interaction between the \textit{empathic} agent and the \textit{subject} agent. 

\begin{gather*}
    \frac{\neg \texttt{affRelEv}(\textit{te}) \; \wedge \; \textit{tg} =   \textit{\textit{emph\_ag\_id}} \; \wedge \; \textit{sbj} \notin \left\{\texttt{null},\texttt{self}\right\} }{\left \langle \textit{ag},C,M,T,s,\textit{Mem},\textit{Ta},\textit{O},\texttt{EvClass} \right \rangle \rightarrow \left \langle \textit{ag},C,M,T,s,\textit{Mem},\textit{Ta},\textit{O}',\texttt{EvClass} \right \rangle} \tag{EvClass$_2$}
\end{gather*}
\begin{align*}
    \textit{where: }  &T_{\varepsilon} = \left \langle \textit{te},i \right \rangle\\
                    &\textit{tg}=\texttt{getTarget}(\textit{te})\\
                    &\textit{sbj} = \texttt{getSubject}(\textit{te})\\
                    &\textit{iv} =  \texttt{getIV}(\textit{te})\\
                    &O' = \left\{...,\left< \textit{id}, \textit{al}', \vec{\sigma} \right>,...\right\}, ~\textit{where}~ \textit{id} = \textit{sbj}\\
                    &\textit{al}' = \texttt{updateAl}(\textit{al},\textit{iv}), ~\textit{where}~ \left< \textit{id}, \textit{al}, \vec{\sigma} \right> \in O ~\text{and}~ \\
                    &\textit{id} = \textit{sbj}
\end{align*}

When this rule is evaluated as true, the affective link that the \textit{empathic} agent has with the \textit{subject} agent is updated by the function \texttt{updateAl}. 

Continuing with the example, suppose that Lily (the \textit{empathic} agent) has an affective link of $0.5$ with Marshall. At a certain point in time, Marshall says hello to Lily by the event:

\begin{gather*}
\begin{aligned}
\texttt{hello[subject(Marshall),target(Lily),}\\
\texttt{interaction\_value(0.2)]} 
\end{aligned}
\tag{Example 1}
\end{gather*}

In this event, Marshall is the \textit{subject} agent. As a result of this interaction, Lily will increase her affective link with Marshall by using the interaction value $0.2$. As in the previous rule, Lily will remain in the \textit{event classification } step waiting for a new event.

The last two rules refer to events that are affectively relevant and, therefore, will trigger the affective cycle. It is in these rules where the distinction between self and \textit{target} occurs. This distinction is crucial because it allows an agent to differentiate its own emotions from emotions produced by the perception of others. The EvClass$_3$ rule is evaluated as true when affective events are directed towards other agents:

\begin{gather*}
    \frac{\texttt{affRelEv}(\textit{te})  \; \wedge \; \textit{tg} \notin  \left \{ \text{\textit{emph\_ag\_id}},\texttt{null} \right \} }{\left \langle \textit{ag},C,M,T,s,\textit{Mem},\textit{Ta},\textit{O},\texttt{EvClass} \right \rangle \rightarrow \left \langle \textit{ag},C,M,T,s,\textit{Mem},\textit{Ta},\textit{O}',\texttt{EmphAppr} \right \rangle} \tag{EvClass$_3$}
\end{gather*}
\begin{align*}
    \textit{where: }  &T_{\varepsilon} = \left \langle \textit{te},i \right \rangle\\
                    &\textit{tg}=\texttt{getTarget}(\textit{te})\\
                    &\textit{iv} = \texttt{getIV}(\textit{te})\\
                    &O' = \left\{...,\left< \textit{id}, \textit{al}', \vec{\sigma} \right>,...\right\}, ~\textit{where}~ \textit{id} = \textit{tg}\\
                &\textit{al}' = \texttt{updateAl}(\textit{al},\textit{iv}), ~\textit{where}~ \left< \textit{id}, \textit{al}, \vec{\sigma} \right> \in O ~\text{and}~ \\
                &\textit{id} = \textit{tg}
\end{align*}

If an event is affectively relevant and the \textit{target} agent is another agent, the event is considered an empathic event. Therefore, the affective cycle is started, and the \textit{empathic appraisal} step processes the empathic event. For example, when Lily receives the event: 

\begin{align*}
&\texttt{slap} \texttt{[subject(Marshall),target(Barney),}\\ &\texttt{affective\_relevant,interaction\_value(-0.5)]}
\tag{Example 2}
\end{align*}

The function \texttt{affRelEv} evaluates the event as affectively relevant and identifies that the \textit{target} of the triggering event is Barney. Therefore, the affective cycle of agent Lily moves to the appraisal (\texttt{Appr}) step in which the event is evaluated by means of an empathic appraisal. In this appraisal step, Lily infers how Barney feels, and she generates an empathic emotion directed towards Barney. 

Finally, the rule EvClass$_4$ deals with affectively relevant events that are received by the \textit{empathic} agent:

\begin{gather*}
    \frac{\texttt{affRelEv}(\textit{te})  \; \wedge \; \textit{tg} =  \textit{emph\_ag\_id}}{\left \langle \textit{ag},C,M,T,s,\textit{Mem},\textit{Ta},\textit{O},\texttt{EvClass} \right \rangle \rightarrow \left \langle \textit{ag},C,M,T,s,\textit{Mem},\textit{Ta},\textit{O},\texttt{Appr} \right \rangle} \tag{EvClass$_4$}
\end{gather*}
\begin{align*}
    \textit{where: }  &T_{\varepsilon} = \left \langle \textit{te},i \right \rangle\\
                    &\textit{tg}=\texttt{getTarget}(\textit{te})
\end{align*}

When this rule is evaluated as true, the \textit{empathic} agent performs a self-appraisal process of the event obtaining an emotion as a result. Following the previous example, when Lily receives the event:

\begin{align*}
&\texttt{slap}\texttt{[subject(Marshall),target(Lily),}\\ &\texttt{affective\_relevant,interaction\_value(-0.5)]}
\tag{Example 3}
\end{align*}

this event is directed toward Lily (i.e., Lily is the \textit{target} agent). Therefore, Lily will evaluate this event through a self-appraisal process in the \textit{appraisal} (\texttt{Appr}) step. 

\subsubsection{Appraisal}
In the \textit{appraisal} step, the values of the appraisal variables $\textit{Av}$ for the event $\varepsilon$ are determined by the self-appraisal process defined by the function:

\begin{gather}
\begin{aligned}
\texttt{Appraisal}&(\varepsilon ,\textit{bs},\textit{cc},\textit{Av}, \textit{Mem}, \textit{Ap}) := \{ \left [  (x_\textit{id},x_v) \right ]:x_v =  \\
 &\texttt{DeriveAV}(x_\textit{id},\varepsilon ,\textit{bs},\textit{cc}, \textit{Mem}, \textit{Ap}),\forall (x_\textit{id},x_v)\in \textit{Av} \}
\end{aligned}
\label{eq:appr}
\end{gather}

where ${\textit{bs}}$ are the agent's beliefs, ${\textit{cc}}$ are the agent's concerns,  $\textit{Av}$ is a list of tuples $(\textit{identifier}, \textit{value})$ containing the set of appraisal variables, \textit{Mem} is the affective memory, ${\textit{Ap}}$ are the applicable plans, $x_\textit{id}$ is the identifier of an appraisal variable, $x_v$ is the value of the appraisal variable, and \texttt{DeriveAV} is a function that returns the value of each appraisal variable defined as:

\begin{gather}
\begin{aligned}
\texttt{DeriveAV}&(\textit{id},\varepsilon ,\textit{bs},\textit{cc}, \textit{Mem}, \textit{Ap}) :=\\
&[\textit{id},\varepsilon ,\textit{bs},\textit{cc}, \textit{Mem}, \textit{Ap}] \mapsto \lambda
\end{aligned}
\label{eq:DeriveAV}
\end{gather}

where $\lambda$ represents the value of the appraisal variable. Note that this function calculates a different value for each appraisal variable. 

To determine the derived emotions from the appraisal variables, we have defined the \texttt{DeriveEm} function as:

\begin{gather}
\begin{aligned}
\texttt{DeriveEm}&(\textit{Av}):= \textit{Av} \mapsto \textit{Ae} 
\end{aligned}
\label{eq:DeriveEm}
\end{gather}

where \textit{Av} represents the set of appraisal variables and \textit{Ae} represents the set of appraised emotions defined as a set of vectors. Therefore, this function establishes a mapping between the set of appraisal variables and emotions. 

The transition rule for the \textit{appraisal} step of the affective cycle has been defined as:

\begin{gather*}
    \frac{\textit{True}}{\left \langle \textit{ag},C,M,T,s,\textit{Mem},\textit{Ta},\textit{O},\texttt{Appr} \right \rangle \rightarrow \left \langle \textit{ag},C,M,T,s,\textit{Mem},\textit{Ta}',\textit{O},\texttt{EmReg} \right \rangle}  \tag{Appr$_1$}
\end{gather*}
\begin{align*}
    \textit{where: }  &\textit{Ta}'_{\textit{Av}} = \texttt{Appraisal}(T_\varepsilon ,\textit{ag}_{\textit{bs}},\textit{ag}_{\textit{cc}}, \textit{Ta}_\textit{Av}, \textit{Mem}, T_{\textit{Ap}}) \\
                    &\textit{Ta}'_{\textit{Ae}} = \texttt{DeriveEm}(\textit{Ta}_{\textit{Av}}')
\end{align*}

For instance, continuing with the event of Example 3, Lily appraises the \texttt{slap} event eliciting the emotion \textit{sadness} with high intensity.  

Once the appraisal finishes, the next step in the affective cycle is the \textit{emotion regulation} (\texttt{EmReg}) step.

\subsubsection{Emotion regulation}
The \textit{emotion regulation} step adapts the appraised emotion based on different factors such as the mood or personality of the agent. This regulation process affects the probability and intensity of emotions. For example, if the affective link between two agents is high, the probability and intensity of emotions should be increased. The process of \textit{emotion regulation} is established by the \texttt{EmphRegulation} function defined as:

\begin{equation}
\begin{aligned}
    \texttt{EmRegulation}&(\textit{tr},\omega,\vec{\sigma},\textit{Ae} ):=\\
    &\left \{ \left [  \vec{e}'\right ] :\vec{e}' = \varphi_1\left ( \textit{tr},\omega,\vec{\sigma}, \vec{e} \right ) ~ \forall \vec{e} \in \textit{Ae} \right \} 
\end{aligned}
\label{emotionRegulationFormula}
\end{equation}

where $\textit{tr}$ are the personality traits of the agent, $\omega$ is the personality/emotion correlation matrix, $\vec{\sigma}$ is the agent's current mood, \textit{Ae} is the set of appraised emotions and $\varphi_1$ is a function that adapts the emotion's vector $\vec{e}$ to the regulation process. 

Thanks to this regulation of the emotion process, each agent can face an event in a different way depending on his/her personality or mood. 

The transition rule for this step is defined as follows:

\begin{gather*}
    \frac{true}{\left \langle \textit{ag},C,M,T,s,\textit{Mem},\textit{Ta},\textit{O},\texttt{EmReg} \right \rangle \rightarrow \left \langle 
    \textit{ag},C,M,T,s,\textit{Mem},\textit{Ta}',\textit{O},\texttt{EmSel} \right \rangle} \tag{EmSel$_1$}
\end{gather*}
\begin{align*}
 \textit{where: } \textit{Ta}'_{\textit{Ae}} = \texttt{EmRegulation}(\textit{ag}_{P},\textit{Ta}_{\vec{\sigma}},\textit{Ta}_\textit{Ae} )
\end{align*}

Following the previous example in which Lily appraised the \textit{sadness} emotion with high intensity, if Lily's personality has a high component of extraversion, she is more prone to positive emotions. Therefore, the \textit{emotion regulation} process will reduce the intensity of the emotion, producing the emotion \textit{sadness} with medium intensity.  

Once the regulation of the possible emotions has been performed, the next step of the non-empathic affective cycle will be the \textit{emotion selection} (\texttt{EmSel}) step.

\subsubsection{Empathic appraisal}

When the \textit{target} of an affective event is another agent, the event is appraised by the \textit{empathic appraisal} process. This \textit{empathic appraisal} process is a self-projection appraisal that allows an \textit{empathic} agent to understand another agent's emotion or situation. Similarly to the previously defined self-appraisal step, the \textit{empathic appraisal} is defined by the formula:

\begin{gather}
\begin{aligned}
&\texttt{EmphAppraisal}(\varepsilon ,\textit{bs},\textit{cc},\textit{Av}, \textit{Mem}, \textit{Ap}, O) := \{ \left [  (x_\textit{id},x_v) \right ]:\\
& x_v =  \texttt{DeriveAV}(x_\textit{id},\varepsilon ,\textit{bs},\textit{cc}, \textit{Mem}, \textit{Ap}, O),~\forall (x_\textit{id},x_v)\in \textit{Av}  \}
\end{aligned}
\end{gather}

where ${\textit{bs}}$ is the set of beliefs, ${\textit{cc}}$ is the set of concerns, \textit{Mem} is the affective memory, ${\textit{Ap}}$ are the applicable plans, $x_\textit{id}$ is the identifier of an appraisal variable, $x_v$ is the value of an appraisal variable, $\textit{Av}$ is a list of tuples $(\textit{identifier}, \textit{value})$ one for each appraisal variable, $O$ is the \textit{target} information, and \texttt{DeriveAV} follows the definition of Equation~\ref{eq:DeriveAV}. that returns the value of each appraisal variable. Note that the beliefs and concerns may be either those of the empathic agent or those of the \textit{target} agent. In addition, both the number and type of appraisal variables and the function \texttt{DeriveAV} of the empathic process can be different from the self-appraisal process. Empathic emotions can also be distinct from self-appraisal and are elicited through function \texttt{DeriveEmphEm}:

\begin{gather}
\begin{aligned}
\texttt{DeriveEmphEm}&(\textit{Av}):=\textit{Av} \mapsto \textit{Ee}
\end{aligned}
\label{eq:DeriveEmphEm}
\end{gather}

where \textit{Av} represents the set of appraisal variables and \textit{Ee} represents the set of empathic emotions defined as a set of vectors.

The transition rule for the \textit{empathic appraisal} step of the affective cycle has been defined as:

\begin{gather*}
    \frac{True}{\left \langle \textit{ag},C,M,T,s,\textit{Mem},\textit{Ta},\textit{O},\texttt{EmphAppr} \right \rangle \rightarrow \left \langle 
    \textit{ag},C,M,T,s,\textit{Mem},\textit{Ta}',\textit{O},\texttt{EmphReg} \right \rangle}  \tag{EmphAppr$_1$}
\end{gather*}
\begin{align*}
 \textit{where: } &\textit{Ta}'_{\textit{Av}} = \texttt{EmphAppraisal}(T_\varepsilon ,\textit{ag}_{\textit{bs}},\textit{ag}_\textit{cc}, \textit{Ta}_\textit{Av},\textit{Mem}, T_{\textit{Ap}}) \\
 &\textit{Ta}'_{\textit{Ee}} = \texttt{DeriveEmphEm}(\textit{Ta}_{\textit{Av}} )
\end{align*}

For example, when Lily perceives the event described in Example 2, Lily evaluates the event \texttt{slap} through a self-projection appraisal and elicits the emotion \textit{sorry for} with high intensity towards Barney.

Once the \textit{empathic appraisal} generates the set of empathic emotions, the \textit{empathic regulation} (\texttt{EmphReg}) process will adapt the empathic emotions using the mood and personality of the empathic agent.

\subsubsection{Empathic regulation}
The empathy regulation step adapts the empathic emotion according to different factors such as the mood $\textit{Ta}_{\vec{\sigma}}$ or personality $\textit{ag}_{P}$ of the agent and the affective link $\textit{O}_\textit{al}$ between the \textit{empathic} agent and the \textit{target} agent. This regulation process allows the empathic response to be personalized to different \textit{target} agents. For example, if the affective link between the \textit{empathic} agent and the \textit{target} agent is high, the probability and intensity of the empathic emotions will be higher. The process of empathic regulation is established by the \texttt{EmphRegulation} function defined as:

\begin{gather}
\begin{aligned}
    &\texttt{EmphRegulation}\left (\textit{tr},\omega,O,\vec{\sigma},\textit{Ee} \right ) := \\
    &\left \{  \left [  \vec{e}'\right ]:\vec{e}' = \varphi_2\left ( \textit{tr},\omega,O,\vec{\sigma},\vec{e} \right )~\forall \vec{e} \in \textit{Ee}\right \}
\end{aligned}
\label{empathicRegulationFormula}
\end{gather}

where $\textit{tr}$ are the personality traits of the agent, $\omega$ is the personality/emotion correlation matrix, $O$ represents the knowledge that the \textit{empathic} agent knows about the \textit{target} agent, $\vec{\sigma}$ is the agent's current mood, and $\varphi_2$ is a function that modifies the emotion vector.

The transition rule for the empathic regulation is defined as:

\begin{gather*}
    \frac{true}{\left \langle \textit{ag},C,M,T,s,\textit{Mem},\textit{Ta},\textit{O},\texttt{EmphReg} \right \rangle \rightarrow \left \langle 
    \textit{ag},C,M,T,s,\textit{Mem},\textit{Ta}',\textit{O}',\texttt{EmSel} \right \rangle}  \tag{EmphReg$_1$}
\end{gather*}
\begin{align*}
 \textit{where: } 
                &\textit{Ta}'_{\textit{Ee}} = \texttt{EmphRegulation}(T_\varepsilon,\textit{ag}_{P_{\textit{tr}}},\textit{O},\textit{Ta}_{\vec{\sigma}},\textit{Ta}_\textit{Ee} )\\
                &T_{\varepsilon} = \left \langle \textit{te},i \right \rangle\\
                &\textit{iv} = \texttt{getIV}(\textit{te})\\
                &O' = \left\{...,\left< \textit{id}, \textit{al}', \vec{\sigma} \right>,...\right\}, ~\textit{where}~ \textit{id} = \textit{tg}\\
                &\textit{al}' = \texttt{updateAl}(\textit{al},\textit{iv}), ~\textit{where}~ \left< \textit{id}, \textit{al}, \vec{\sigma} \right> \in O ~\text{and}~ \\
                &\textit{tg}=\texttt{getTarget}(\textit{te})
\end{align*}

Following the previous example, let us suppose that Lily has a low positive affective link with Barney (e.g., $0.3$). Let us recall that Lily's personality makes her prone to positive emotions. In addition, when Lily perceives the event, she has a positive mood (e.g., \textit{happiness}). The \textit{empathic emotion regulation} process will reduce the intensity of the empathic emotion and the result will be the emotion \textit{sorry for} with \textit{weak} intensity.  

Once the regulation of the possible empathic emotions has been performed, the following step is the \textit{emotion selection} (\texttt{EmSel}) step.

\subsubsection{Emotion selection}
In the emotion selection step of the affective cycle, the final emotion ${Fe}$ that the agent is going to simulate is selected by the function $\texttt{SelEmotion}(\varepsilon,\textit{Ae},\textit{Ee}, P)$. This function has as parameters an event $\varepsilon$, the list of appraised emotions ${Ae}$, the list of empathic emotions ${Ee}$, and the agent personality ${P}$. The user can adapt this $\texttt{SelEmotion}$ function to different environments and situations considering the agent personality, intensity or probability of the emotions, and other affective characteristics.

The transition rule for the \textit{emotion selection} step is defined as follows:

\begin{gather*}
    \frac{True}{\left \langle ag,C,M,T,s,Mem,Ta,\textit{O},\texttt{EmSel} \right \rangle \rightarrow \left \langle ag,C,M,T,s,\textit{Mem}',Ta',\textit{O},\texttt{AffAd} \right \rangle} \tag{EmSel$_1$}
\end{gather*}
\begin{align*}
 \textit{where: } & \textit{Ta}'_{\textit{Fe}} = \texttt{SelEmotion}(\textit{Ta}_{\textit{Ae}},\textit{Ta}_{\textit{Ee}})\\
                &\textit{ae} = \left \langle T_\varepsilon, \textit{Ta}_\textit{Fe}' \right \rangle\\
                &\textit{Mem}' = \textit{Mem} \cup \textit{ae}
\end{align*}

This step calculates the final emotion $\textit{Ta}'_{\textit{Fe}}$. Therefore, the affective memory \textit{Mem} can be updated to store the event $T_\varepsilon$ with the emotion $\textit{Ta}'_{\textit{Fe}}$ resulting from this event appraisal. This mechanism allows the agent to maintain a memory of past events and the emotion that each event produced, supporting the maintaining of long-term interactions between agents. 

For instance, when Lily evaluates the event of Example 3, the self-appraisal process produces two emotions as a result: \textit{sadness} with high intensity, and \textit{fear} with \textit{medium} intensity. Let us suppose that after the \textit{emotion regulation} process the intensity of those emotions is not modified. If the user has defined the $\texttt{SelEmotion}$ function to select the emotion with the highest intensity, the empathic agent will elicit the \textit{sadness} emotion.   

The next process of the empathic affective cycle is the \textit{affect adaptation} (\texttt{AffAd}).

\subsubsection{Affect adaptation}

In the \texttt{AffectAdaptation} step, the mood of the agent is updated by the function \texttt{AffectAdaptation}, which is defined as follows: 

\begin{gather}
\begin{aligned}
    \texttt{AffectAdaptation}\left (\textit{tr},\omega,\vec{\sigma},\textit{Fe} \right ) := \left\{  \vec{\sigma}':\vec{\sigma}' = \varphi_3\left ( \textit{tr},\omega,\textit{Fe},\vec{\sigma} \right )\right\}
\end{aligned}
\label{affectadaptationForumla}
\end{gather}

where \textit{Fe} is the final emotion selected by the \textit{emotion selection} process. $\varphi_3$ is a function that, similarly to $\varphi_1$ and $\varphi_2$ in Equations~\ref{emotionRegulationFormula}~and~\ref{empathicRegulationFormula}, modifies the mood vector $\vec{\sigma}$ to adjust it to the emotion produced by the event.

The transition rule for this step is defined as:

\begin{gather*}
     \frac{True}{\langle \textit{ag},C,M,T,s,\textit{Mem},\textit{Ta},\textit{O},\text{\texttt{AffAd}} \rangle \to \langle \textit{ag},C,M,T,s, \textit{Mem},\textit{Ta}', P,\textit{O},\text{\texttt{SelCs}} \rangle} \tag{AffAd$_1$}
\end{gather*}
\begin{align*}
 \textit{where: } \textit{Ta}'_{\vec{\sigma}} = \texttt{affectAdaptation}(\textit{ag}_P,\textit{Ta}_{\vec{\sigma}},\textit{Ta}_\textit{Fe})
\end{align*} 

Since emotions and mood are expressed as vectors in the same representation space, the selected emotion will ``attract'' the mood to a greater or lesser extent depending on the agent’s personality. As a result, the mood of the agent will be modified. Continuing with the previous example in which the selected emotion was sadness with high intensity, let us assume that Lily's mood is \textit{happiness} with a low intensity. The emotion will attract the mood, and, as a result, the new mood will be \textit{sadness} with a \textit{low} intensity.

Once the mood of the agent is updated, the next state of the affective cycle will be the \textit{select coping strategy} process. In this process, a coping strategy from the current set of applicable coping strategies will be used to create a plan that finally allows the agent to express its empathy: in its voice tone, in its speech, in its face, \ldots

\section{Default design}\label{defaultDsign}

To facilitate the use of \egenia, we propose a default design in which there is a default definition of the previous presented functions, based on the psychological theories introduced in Sections~\ref{section:genia} and~\ref{proposal}. In this default design, emotions and mood are represented as vectors $\vec{\sigma}$ in a representation space based on the PAD model \cite{mehrabian1996pleasure}. The personality is defined by the OCEAN model. The appraisal model is based on the OCC model \cite{ortony1990cognitive} and follows the definition presented in \cite{alfonso2017toward}. The appraisal function evaluates the event, when this event implies the addition or deletion of a belief. Then the values of the appraisal variables are estimated considering both the information contained in the event itself and the agent's beliefs, desires, intentions, expectations, and concerns. This default design uses the appraisal variables: \textit{desirability}, which considers the value of the agent's concerns about the event; \textit{likelihood}, which considers the probability of the event; and \textit{causal attribution}, which is the agent that produces the event (i.e. the subject agent) but it can also be \textit{self}, if it is an internal event or \textit{null} if it comes from the environment (see \cite{alfonso2017toward} and \cite{tavernerIS} for more details on how this process work). The function \texttt{DeriveAV} in Equation~\ref{eq:DeriveAV} estimates the value of the appraisal variables as follow. For the \textit{desirability}, this function evaluates if the event is an addition or deletion event, then compares the functor of the event with the agent concerns and estimates the desirability value. For example, if the agent is concerned about passing an exam, the function for estimating the value of the concern may be $V = \textit{Score}/\textit{MaxScore}$, where \textit{Score} is the score of the agent's exam and \textit{MaxScore} is the maximum possible score. To obtain the value of the appraisal variable \textit{likelihood}, a special annotation $\textit{prob\_\_}(n)$ is used for events. Finally, to obtain the value for the \textit{causal attribution}, the Equation~\ref{eq:getSubject}, referring the function \texttt{getSubject}, is used. In our default design, \texttt{EmphAppraisal} and \texttt{DeriveEmphEm} (Equation~\ref{eq:DeriveEmphEm})  functions follow the same philosophy as \texttt{Appraisal} and \texttt{DeriveEm} functions (Equations~\ref{eq:appr} and \ref{eq:DeriveEm}) presented above. 
Note that, in this function, the appraised beliefs and concerns to be those of the agent itself or of the \textit{target} agent allowing the simulation of different levels of empathy and Theory of Mind. In addition, the number and type of emotions that can be triggered by empathic appraisal, may differ from the emotions of affective appraisal. For example, by triggering emotions such as \textit{"happy for"} or \textit{"sorry for"} proposed in the OCC model \cite{ortony1990cognitive}. These two emotions are used in the default design. 

In our default design, the \texttt{DeriveEm} function, Equation~\ref{eq:DeriveEm}, can elicit five emotions (i.e., hope, joy, fear, sadness, and guilt) using the OCC theory as proposed in \cite{marsella2009ema}. These emotions can be subsequently represented in a PAD space, considering the mapping proposed in \cite{gebhard2005alma} (or to a \textit{Pleasure-Arosual} (PA) space, considering the model proposed in \cite{taverner2020multidimensional}). For more details see \cite{alfonso2017toward}.

Once the emotion is obtained and represented as a vector $\vec{\sigma}$, the emotion regulation process adapts this vector $\vec{\sigma}$ through the function $\varphi_1$ (Equation~\ref{emotionRegulationFormula}). This function is defined as: 

\begin{equation}
\varphi_1\left ( \textit{tr},\omega,\vec{\sigma}, \vec{e} \right ) := \psi(\vec{\sigma}) \cdot \vec{\sigma} + \vec{e}
\end{equation}

where $\psi(\vec{v})$ is a weighting factor resulting from the formula:

\begin{equation}
 \psi(\vec{v}) :=  \frac{\sum_{i \in \textit{tr}} \beta_i \cdot \omega_{t(\vec{v}),i}}{\sum_{j \in \textit{tr}} \omega_{t(\vec{v}),j}}
\end{equation}

where $\beta_i$ is the value of the $i$th personality trait, $t(\vec{v})$ is a function that returns the most probable affective label for a vector $\vec{v}$ (the description of function $t(\vec{v})$ can be found at \cite{tavernerIS} and \cite{taverner2020multidimensional}), and $\omega_{t(\vec{\sigma}),i}$ represents the value of the correlation matrix $\omega$ for the affective label $t(\vec{\sigma})$ and the $i$th personality trait.

The empathic regulation function $\varphi_2$, proposed in Ecuation~\ref{empathicRegulationFormula}, is defined in the default design as:

\begin{gather}
\begin{aligned}
    \varphi_2\left (\textit{tr},\omega,O,\vec{\sigma}, \vec{e} \right ) := \left (\psi(\vec{e})  \cdot \vec{e} + \psi(\vec{\sigma}) \cdot \vec{\sigma} \right ) \cdot \textit{al}
\end{aligned}
\end{gather}

where \textit{al} represents the affective link. 

The emotion selection (\texttt{SelEmotion}) function is based on the intensity and probability of emotions according to the following definition:

\begin{equation}
\texttt{SelEmotion}(\textit{Ae},\textit{Ee}) := \underset{i\in\textit{Ae}\cup\textit{Ee}}{\arg\max}\left (  \underset{c}{\arg\max}\, \widehat{P} (C=c \mid \alpha_i)\cdot\delta_i\right )
\end{equation}

where $\alpha_i$ and $\delta_i$ represent the angle and the intensity of the $i$th emotion vector (as defined in \cite{taverner2020multidimensional}), respectively. 

For the affect adaptation, Equation~\ref{affectadaptationForumla}, the definition in the default design is:

\begin{gather}
\begin{aligned}
    \varphi_3\left ( \textit{tr},\omega,\textit{Fe},\vec{\sigma} \right ) := \left\{ \vec{\sigma} \cdot \psi(\vec{\sigma}) + \vec{e} \cdot \psi(\vec{e}) \right\}
\end{aligned}
\end{gather}

Finally, the function for updating the affective link (\texttt{updateAl}) is defined as:

\begin{gather}
\texttt{updateAl}(\textit{al},\textit{iv}) :=  \textit{al} + \varphi \cdot \textit{iv} 
\end{gather}

where \textit{iv} is the interaction value, $\varphi$ is a weighting factor, and \textit{al} and represent the current value of the affective link. 

\section{Conclusions}\label{conclusion}
Empathy is a key element in social interactions affecting human affective states and behaviours and provides a basis for long-term relationships. Several proposals have been made from the area of affective computing to simulate affective and empathic abilities. These proposals are generally based on theoretical models from psychology, sociology, and neuroscience using agent-oriented approaches. However, these proposals are designed ad hoc, generating agent programs that are highly dependent on the domain and the programmer, making these proposals very difficult to understand, share, and re-use. In this paper, we have presented {\egenia} a generic architecture for the development of empathic agents. We have proposed a formalization of the  syntax, semantics, and the reasoning cycle of AgentSpeak to support the development of agents with empathic abilities. {\egenia} formalizes the three main processes identified in the literature: perception, empathic appraisal, and empathic regulation. After perceiving an empathic event, the empathic appraisal process elicits a set of empathic emotions. These empathic emotions are then adapted by the empathic regulation process according to the knowledge that the agent has about the \textit{target} agent.
We have modified the affective cycle that was incorporated to AgentSpeak by \genia adding new processes and functionalities to allow the elicitation of empathic emotions. We have formalized our extension using the same operational semantics formalism used by AgentSpeak and {\genia}. This operational semantics defines the structure and the configuration of the agent allowing the use of different psychological theories. The formalization of all of the transitions guarantees the validity of our proposal. 

The new affective cycle is divided into two different appraisal processes: one for eliciting empathic emotions induced by the perception of an emotion or situation in a \textit{target} agent; and another for eliciting emotions based on self-directed or environmental events. This division of the affective process allows the agent to be able to maintain a sense of self that is distinct from the \textit{target} agent.

The use of the affective link representing the affective proximity or social tie between two agents allows the personalization of the empathic response based on the \textit{target} agent. In addition, a correlation matrix represents how each personality trait of the agent influences the emotion that will be elicited. Finally, in our proposal, agents maintain a memory of past events and the emotion that the past events produced in the agent. This affective memory, combined with the possibility of establishing and modifying the affective links with other agents, allows the simulation of personalized long-term interactions, which are crucial in affective interactions in human beings. All of these characteristics of our empathic agent model will produce software agents that are more realistic when modelling human organizations.

We are currently considering increasing the agent's knowledge of other agents by adding new structures to store the beliefs, goals, and concerns of other agents using state-of-the-art approaches of Theory of Mind. Moreover, the affective link can be complemented maintaining a level of trust in other agents. This will improve the perspective-taking process so that the agent can evaluate perceived situations using more knowledge about the \textit{target} agent.

\section*{Acknowledgments}
\noindent This work is partially supported by the Spanish Government project PID2020-113416RB-I00 and TAILOR, a project funded by EU Horizon 2020, under GA No 952215.

\bibliographystyle{plain}
\bibliography{bibliography}

\end{document}